\documentclass{article}
\usepackage{spconf,amsmath,amsfonts,graphicx}
\usepackage{multirow}
\usepackage{hyperref}

\name{Hieu-Thi Luong\textsuperscript{1}, Junichi Yamagishi\textsuperscript{1,2}\thanks{This work was partially supported by MEXT KAKENHI Grants (16H06302, 17H04687, 18H04120, and 18H04112).}}
\address{\textsuperscript{1}National Institute of Informatics, Tokyo, Japan\quad{}\quad{}\textsuperscript{2}The University of Edinburgh, Edinburgh, UK}
\title{Scaling and bias codes for modeling speaker-adaptive \\ dnn-based speech synthesis systems}

\begin{document}

\maketitle

\begin{abstract}
Most neural-network based speaker-adaptive acoustic models for speech synthesis can be categorized into either layer-based or input-code approaches. Although both approaches have their own pros and cons, most existing works on speaker adaptation focus on improving one or the other. In this paper, after we first systematically overview the common principles of neural-network based speaker-adaptive models, we show that these approaches can be represented in a unified framework and can be generalized further. More specifically, we introduce the use of scaling and bias codes as generalized means for speaker-adaptive transformation. By utilizing these codes, we can create a more efficient factorized speaker-adaptive model and capture advantages of both approaches while reducing their disadvantages. The experiments show that the proposed method can improve the performance of speaker adaptation compared with speaker adaptation based on the conventional input code. 
\end{abstract}
%
\section{Introduction}
\label{sec:intro}

Recent speaker-dependent speech synthesis systems can generate high-quality reading speech indistinguishable from natural human speech when their training data is recorded in a quality-controlled condition and have sufficient amount of data \cite{shen2017natural}. The speech synthesis community is currently trying to solve more challenging problems. A good example is multi-speaker speech synthesis and its adaptation \cite{parker2018adaptation,wang2018style,lorenzo2018investigating,inoue2017investigation}. Here multi-speaker synthesis means generating synthetic speech of multiple known speakers included in a training dataset using a common model, and adaptation means adapting the speaker-independent common model to unseen speakers and generating their speech. This speaker-adaptive speech synthesis systems are expected to opens possibilities for a wide range of new applications for speech synthesis such as a customizable, user-specific voice interface and voice preservation for people with medical conditions involving voice losses. However, training the multi-speaker synthesis models and adapting them to unseen speakers are still challenging problems, and resulting models are far from perfect, especially when less than ideal datasets are used \cite{jia2018transfer}. 

Most adaptation methods for neural network models can be described as either (a) fine-tuning a set of or all of parameters of speaker-independent network so it explains unseen speaker's data better or (b) factorizing a neural network into speaker-specific and common parts and estimating the speaker-specific components for the unseen speaker's data. The speaker-specific components may be composed by input codes (e.g.\ one-hot vector) \cite{luong2017adapting}, embedding vectors obtained externally  (e.g.\ i-vector) \cite{wu2015study}, or latent variables (e.g.\ variational auto-encoder) \cite{wang2018style,akuzawa2018expressive,delcroix2018auxiliary}. Of course any of those speaker-specific components may be jointly optimized with the common parts (e.g.\ \cite{luong2017adapting,delcroix2018auxiliary,wan2017integrated}).
Although there are a lot of variants on multi-speaker modeling and adaptation, most approaches for augmenting the speaker-specific components into a neural network are equivalent to adapting a bias term of each hidden layer and this bias term is typically constant across all frames of all utterances. Although Wu et al. \cite{wu2018feature} and Nachmachi et al. \cite{nachmani2018fitting} proposed frame-dependent components, these components are still bias adaptation and their underlying frameworks and concepts have mathematical similarities. 

In this paper we first systematically overview the common concepts of neural-network based speaker-adaptive models and show that these approaches can be represented in a unified framework. Further, we introduce a scaling code as an extended speaker-adaptive transformation. As its name indicates, this code introduces an additional scaling operation as an approximation to  adaptation of weight matrices unlike the conventional deep neural network (DNN) adaptation approaches.
Section \ref{sec:background} details relevant work. Section \ref{sec:architecture} describes our factorized speaker adaptation based on  scaling and bias codes. Section \ref{sec:exp} explains our experiments and shows both objective and subjective results. We conclude our work and describe the future direction for this method in Section \ref{sec:conclusions}.

\section{Related work}
\label{sec:background}

Constrained Maximum Likelihood Linear Regression (CMLLR) \cite{digalakis1995speaker,gales1998maximum}, also known as feature-space MLLR (fMLLR), is a widely used speaker adaptation technique for hidden Markov model (HMM)-based speech processing systems in which a speaker-dependent affine transformation is applied to source acoustic features to explain target data better. In the case of automatic speech recognition (ASR), the transformation acts as a method of normalization, whereas in the case of speech synthesis, the transformation purpose is to diverge the acoustic output to each target speaker \cite{yamagishi2009analysis}. The fMLLR method can be described using the following equation:
\begin{equation}
    \overline{\mathbf{x}} = \mathbf{A}^{(k)}\mathbf{x} + \mathbf{b}^{(k)}
\end{equation}
where $\mathbf{x}$ is the source acoustic features, $\overline{\mathbf{x}}$ represents approximated acoustic features of the target speaker $k$, $\mathbf{A}^{(k)}$ is a full linear matrix and $\mathbf{b}^{(k)}$ is the bias vector. $\mathbf{A}^{(k)}$ and $\mathbf{b}^{(k)}$ are transformation parameters specific to each speaker. 

A feedforward layer of a standard neural network can be defined by the following equation:
\begin{equation}
\label{eq:vanillalayer}
    \mathbf{h}_{l} = f( \mathbf{W}_{l}\mathbf{h}_{l-1} + \mathbf{c}_l )
\end{equation}
where $\mathbf{h}_l$ is the output of the $l$-th hidden layer. To simplify our equation, let us assume all hidden layers have the same number of hidden units $m$, that is, $\mathbf{h}_l, \mathbf{h}_{l-1} \in \mathbb{R}^{m \times 1}$ and the $l$-th hidden layer has a weight matrix $\mathbf{W}_l \in \mathbb{R}^{m \times m}$ and a bias vector $\mathbf{c}_l \in \mathbb{R}^{m \times 1}$. $f(.)$ is an element-wise non-linear activation function (such as sigmoid or tanh) that deterministically squashes each dimension of an input vector $\mathbb{R}^{m \times 1}$ to a limited range. 

Next we explain the existing DNN-based speaker adaptation methods, that is, speaker-dependent layers and speaker-dependent input code using similar notations to the above fMLLR.
For the speaker-dependent layers \cite{fan2015multi,huang2018linear} approach, the weight matrices and bias vectors of specific layers are fine-tuned using adaptation data, therefore we can rewrite Equation \ref{eq:vanillalayer} as:
\begin{equation}
\label{eq:fulllayer}
    \overline{\mathbf{h}}_{l} = f( \mathbf{W}_{l}^{(k)}\mathbf{h}_{l-1} + \mathbf{c}_l^{(k)} )
\end{equation}
where $\mathbf{W}_{l}^{(k)}$ and $\mathbf{c}_l^{(k)}$ are now specific to a target speaker $k$ and $\overline{\mathbf{h}}_{l}$ also represents an adapted hidden layer . The method has the advantage of modeling both a full matrix $\mathbf{W}_{l}^{(k)}$ and the bias vector $\mathbf{c}_l^{(k)}$, which usually yield favorable result when the adaptation data is sufficient \cite{wu2015study,huang2018linear}. However when the amount of adaptation data is limited, the result is unstable as number of parameters estimated is very large \cite{hojo2018dnn}. This is also the reason that this method typically involves reducing the number of parameters estimated \cite{samarakoon2016factorized,zhao2017extended,huang2018linear} in order to retain the adaptation performance. 

Learning Hidden Unit Contribution (LHUC) \cite{swietojanski2014learning} is an adaptation method that transforms outputs of the activation function using a speaker-dependent  diagonal transformation matrix, which significantly reduces the number of parameters:
\begin{equation}
\label{eq:lhuc}
    \overline{\mathbf{h}}_{l} = {\rm Diag} \mathbf{A}_l^{(k)} \circ f( \mathbf{W}_{l}\mathbf{h}_{l-1} + \mathbf{c}_l )
\end{equation}
where $\mathbf{A}_l^{(k)} \in \mathbb{R}^{m \times m}$ is a diagonal matrix for speaker $k$, ${\rm Diag}$ is an operation to extract diagonal elements of a ${m \times m}$ matrix as a ${m \times 1}$ vector, and $\circ$ is an element-wise multiplication of vectors. In LHUC, since we apply the transformation after the activation function of the current layer, we may write the LHUC operation at the next hidden layer as follows:
\begin{align}
\label{eq:lhuc2}
    \overline{\mathbf{h}}_{l+1} &= f( \mathbf{W}_{l+1}\overline{\mathbf{h}}_{l} + \mathbf{c}_{l+1} ) \\
                     &= f\left( \mathbf{W}_{l+1} \cdot \left({\rm Diag} \mathbf{A}^{(k)}_l \circ \mathbf{h}_{l} \right) + \mathbf{c}_{l+1} \right) \\
                     &= f\left( \mathbf{W}_{l+1} \mathbf{A}^{(k)}_l \mathbf{h}_{l} + \mathbf{c}_{l+1} \right) \\
                     &= f\left( \mathbf{W}^{(k)}_{l+1} \mathbf{h}_{l} + \mathbf{c}_{l+1} \right) 
\end{align}
From these equations, we see that a speaker-specific weight matrix $\mathbf{W}^{(k)}_{l+1}$  is factorized as $\mathbf{W}_{l+1} \mathbf{A}^{(k)}_l$. 

For the speaker-dependent input-code approach, a vector representing the speaker identity is fed into one or many layers of a neural network. This vector can be as simple as an one-hot vector \cite{luong2017adapting,hojo2018dnn} or an embedding vector obtained from outside systems like speaker verification \cite{jia2018transfer,arik2018neural} or speaker recognition \cite{doddipatla2017speaker}. Although there are many variations, each may be viewed as a bias adaptation of a hidden layer and the speaker-dependent input approach can be written as:
\begin{align}
\label{eq:inputcode}
    \overline{\mathbf{h}}_{l} &= f( \mathbf{W}_{l}\mathbf{h}_{l-1} + \mathbf{c}_l + \mathbf{W}^{b}_l\mathbf{s}^{(k)}) \\
                              &= f( \mathbf{W}_{l}\mathbf{h}_{l-1} + \mathbf{c}_l^{(k)})
\end{align}
where $\mathbf{s}^{(k)} \in \mathbb{R}^{q \times 1}$ is the auxiliary input vector specific to speaker $k$ and has an arbitrary size $q$; $\mathbf{W}^{b}_l \in \mathbb{R}^{m \times q}$ is a new weight matrix added to the layer to handle the new input.  The input code approach provides the flexibility of using an outside system to constrain the model. It is also convenient to present each speaker (or speaking style) as one single vector since it may be used for controlling characteristics of synthetic speech \cite{luong2017adapting,wang2018style,skerry2018towards}. As the number of speaker-dependent parameters $q$ is typically small, this method shows preferable results when the amount of adaptation data is limited. However, it does not seem to improve the adaptation performance when the adaptation data is plentiful \cite{hojo2018dnn}.  


\section{Factorized speaker transformation based on scaling and bias codes}
\label{sec:architecture}
\subsection{Scaling and bias codes}

\begin{figure}[tb]
  \centering
  \includegraphics[width=0.7\columnwidth]{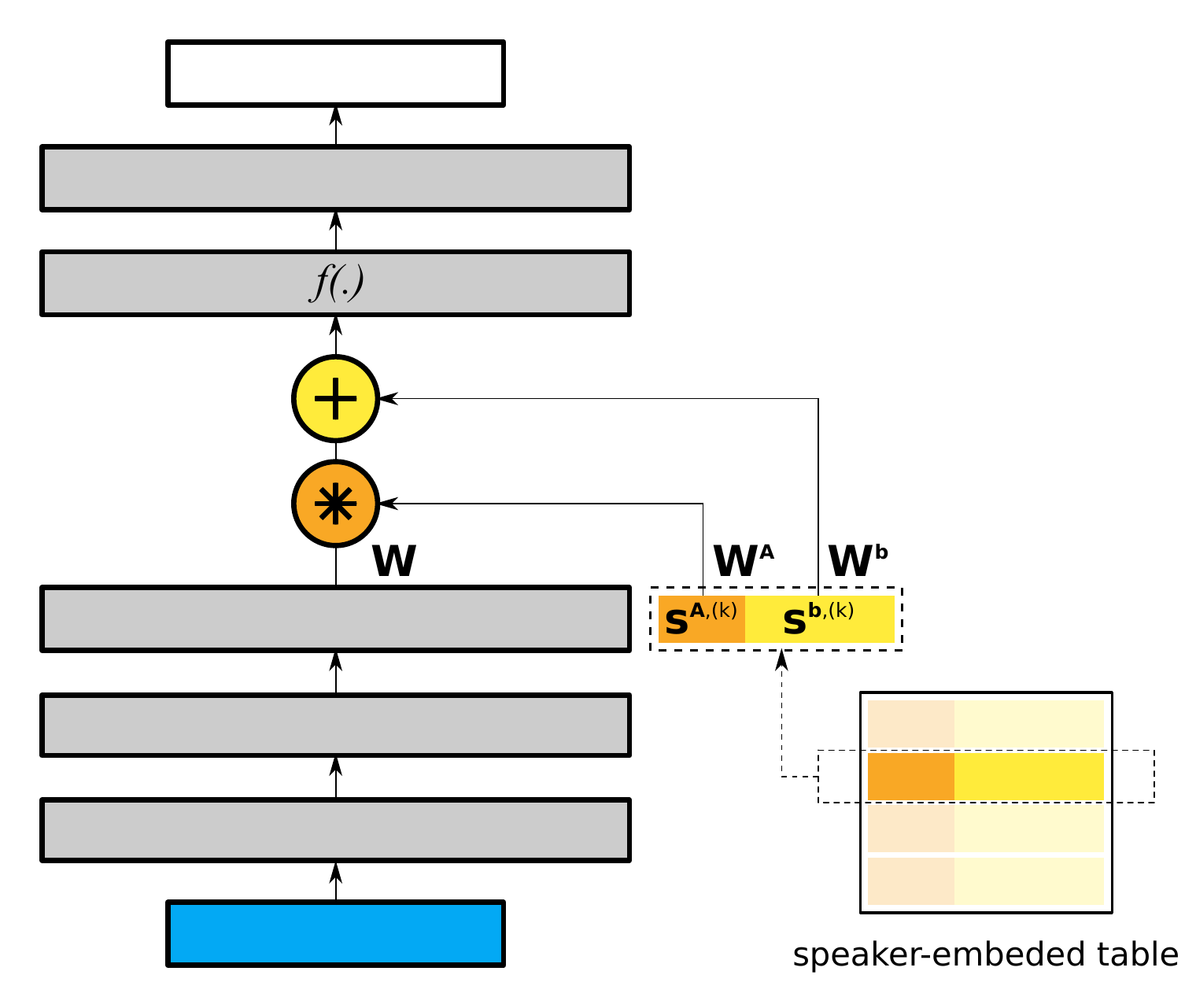}
  \vspace{-3mm}
\caption{Proposed factorized speaker transformation based on scaling and bias codes. Gray boxes indicate layers with nonlinear activation function, and the white box indicates a layer with linear function.}
\label{fig:architecture}
\vspace{-3mm}
\end{figure}

The above approaches are obviously complementary. Our proposal, illustrated in Figure \ref{fig:architecture}, is therefore the design of a new speaker transformation by combining the above two types of approaches and further factorizing its essential components on the basic of ``scaling'' and ``bias'' codes. The main idea is to explicitly transform both the weight matrix and the bias vector as:
\begin{align}
\label{eq:speakerlayer}
    \overline{\mathbf{h}_{l}} &= f( \mathbf{A}^{(k)}_{l} \mathbf{W}_{l} \mathbf{h}_{l-1} + \mathbf{c}_l + \mathbf{b}_l^{(k)} )  \\
\label{eq:scalingcode}
    \mathbf{A}^{(k)}_{l} &= {\rm diag}( \mathbf{W}^{A}_l\mathbf{s}^{A,(k)}) \\
    \mathbf{b}^{(k)}_{l} &= \mathbf{W}^{b}_l\mathbf{s}^{b,(k)}
\end{align}
where $\mathbf{A}^{(k)}_l \in \mathbb{R}^{m \times m}$ is a diagonal matrix for the scaling operation at the $l$-th layer. The matrix is further factorized into a speaker-independent projection matrix $\mathbf{W}^{A}_l \in \mathbb{R}^{m \times p}$ and a scaling code vector $\mathbf{s}^{A,(k)} \in \mathbb{R}^{p \times 1}$. ${\rm diag}$ is an operation to change a ${m \times 1}$ vector into a diagonal ${m \times m}$ matrix. The speaker-specific bias term $\mathbf{b}^{(k)}_{l}$ is also factorized in the same way using $\mathbf{W}^{b}_l \in \mathbb{R}^{m \times q}$ and $\mathbf{s}^{b,(k)} \in \mathbb{R}^{q \times 1}$. As described previously, $\mathbf{s}^{b,(k)}$ is basically equivalent to the conventional speaker code, but we call it as bias code here to better outline its property. These codes may have arbitrary lengths, but, $p$ and $q$ are usually chosen to be much smaller than $m$ to reduce the number of free parameters further.

Factorizing models explicitly and using lower-dimensional subspaces is a powerful concept used in various models (e.g.\ Heteroscedastic Linear Discriminant Analysis (HLDA) \cite{kumar1998heteroscedastic}, subspace Gaussian mixture model \cite{povey2011subspace}). The proposed factorization is somewhat similar to Factorize Hidden Layer (FHL) introduced by Samrakoon and Sim \cite{samarakoon2016factorized}, but we focus on performing the scaling and bias adaptation simultaneously using lower dimensional vectors. A concept similar to scaling and bias codes was also investigated for ASR in \cite{cui2017embedding,samarakoon2017learning}, but instead of mapping the scaling and bias transformation from a common vector we use separated vectors as scaling and bias codes to give ourselves more degrees of freedom to design a speaker-adaptive architecture. If necessary, we may directly adapt $\mathbf{A}^{(k)}_l$ and $\mathbf{b}^{(k)}_l$ when the amount of adaptation data is sufficient.

\subsection{Extensions of the proposed method}

\begin{figure}[tb]
  \centering
  \begin{tabular}{cc}
  \includegraphics[width=0.45\columnwidth]{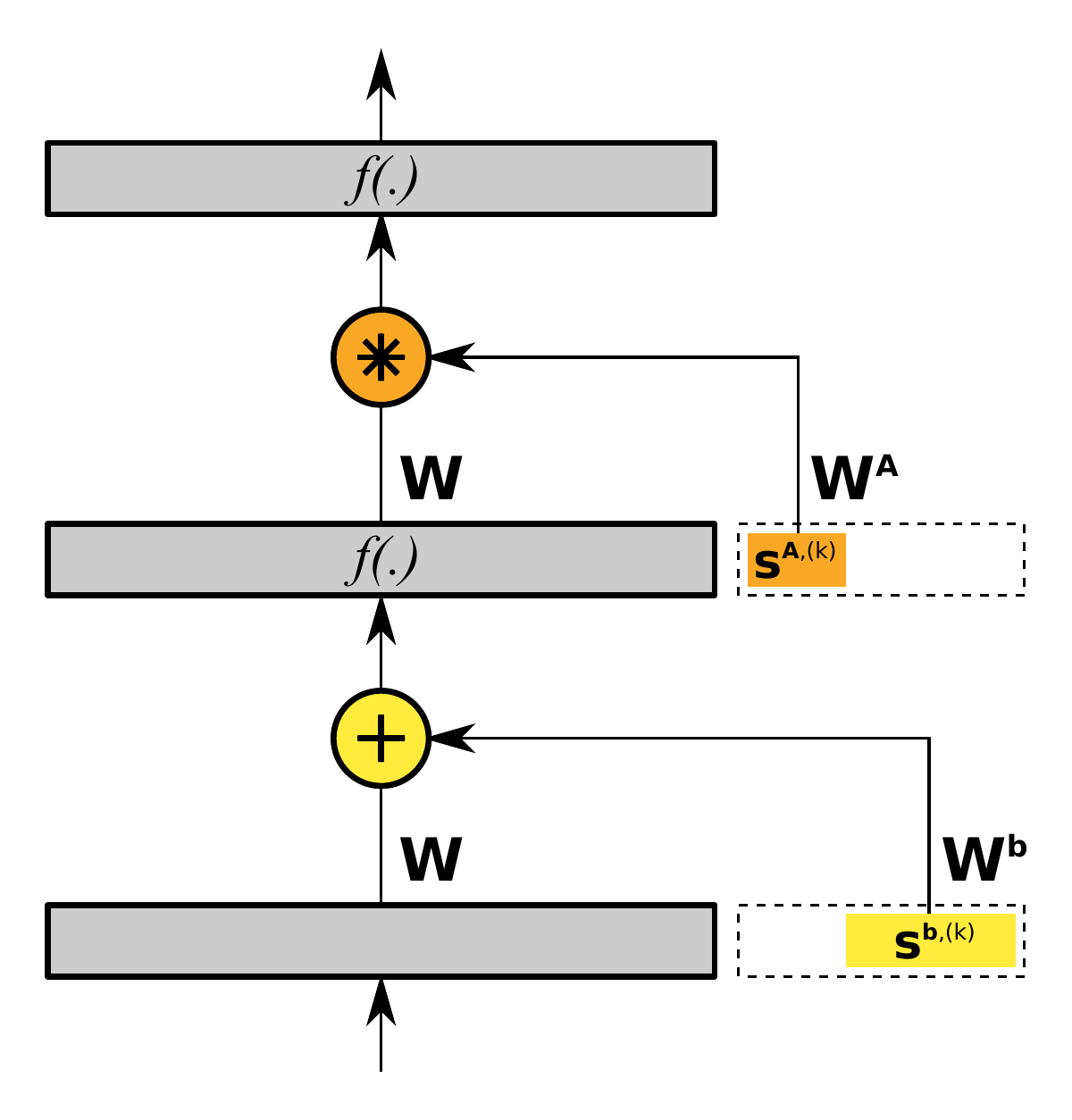} & 
  \includegraphics[width=0.45\columnwidth]{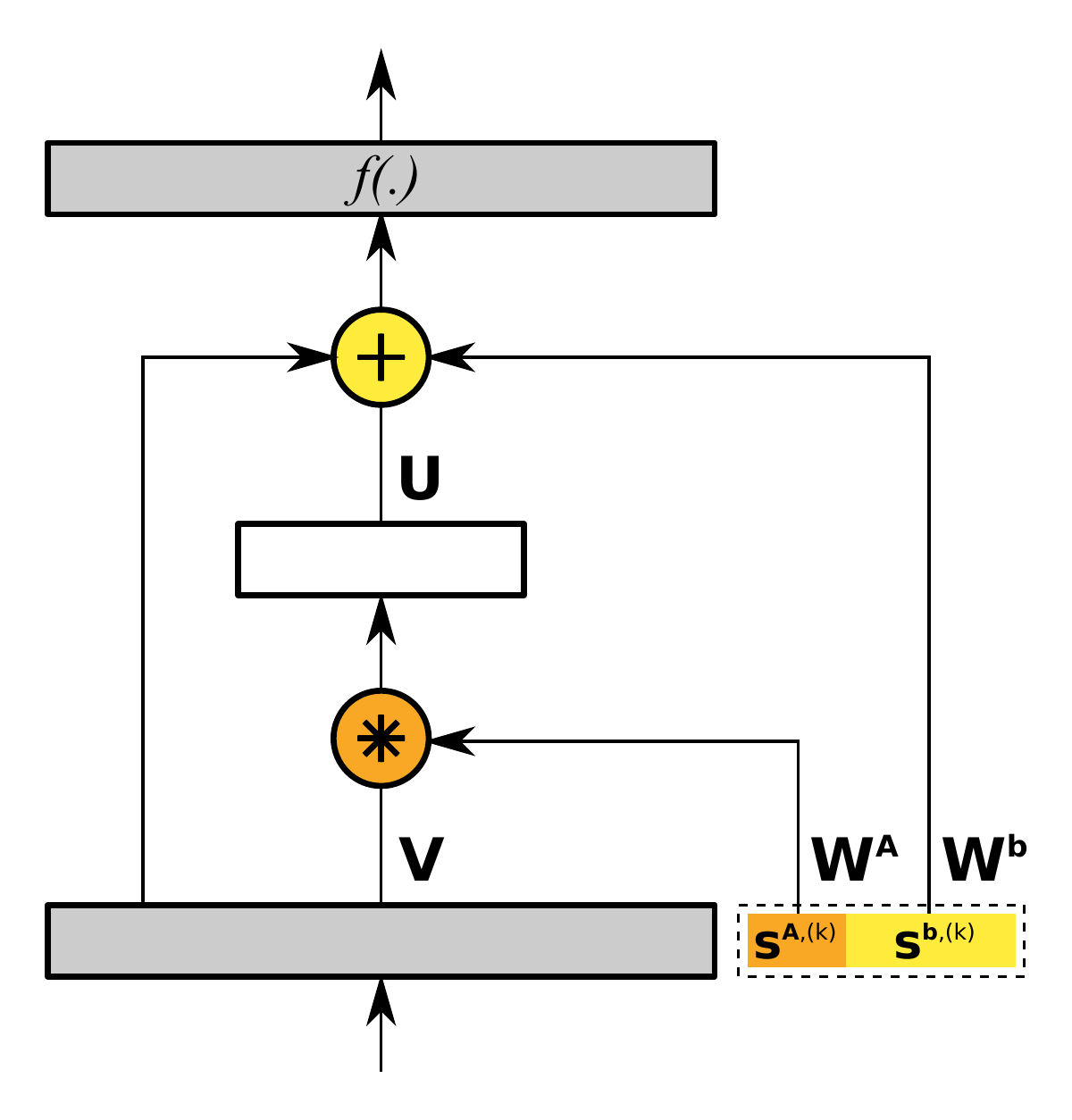} \\
  (a) Multilevel & (b) Bottleneck
  \end{tabular}
  \vspace{-5mm}
\caption{Extended strategies utilizing the scaling and bias codes to integrate speaker transformations into neural network}
\label{fig:strategies}

\end{figure}

In this paper, we investigate two more strategies as extensions of the proposed method. 
The first strategy is to separately use the scaling and bias codes at different layers and to explicitly perform either scaling or bias operations only as illustrated by Figure 
\ref{fig:strategies}-a. 
This is a special case of the proposed method.

\begin{figure}[tb]
  \centering
  \begin{tabular}{cc}
  \includegraphics[width=0.45\columnwidth]{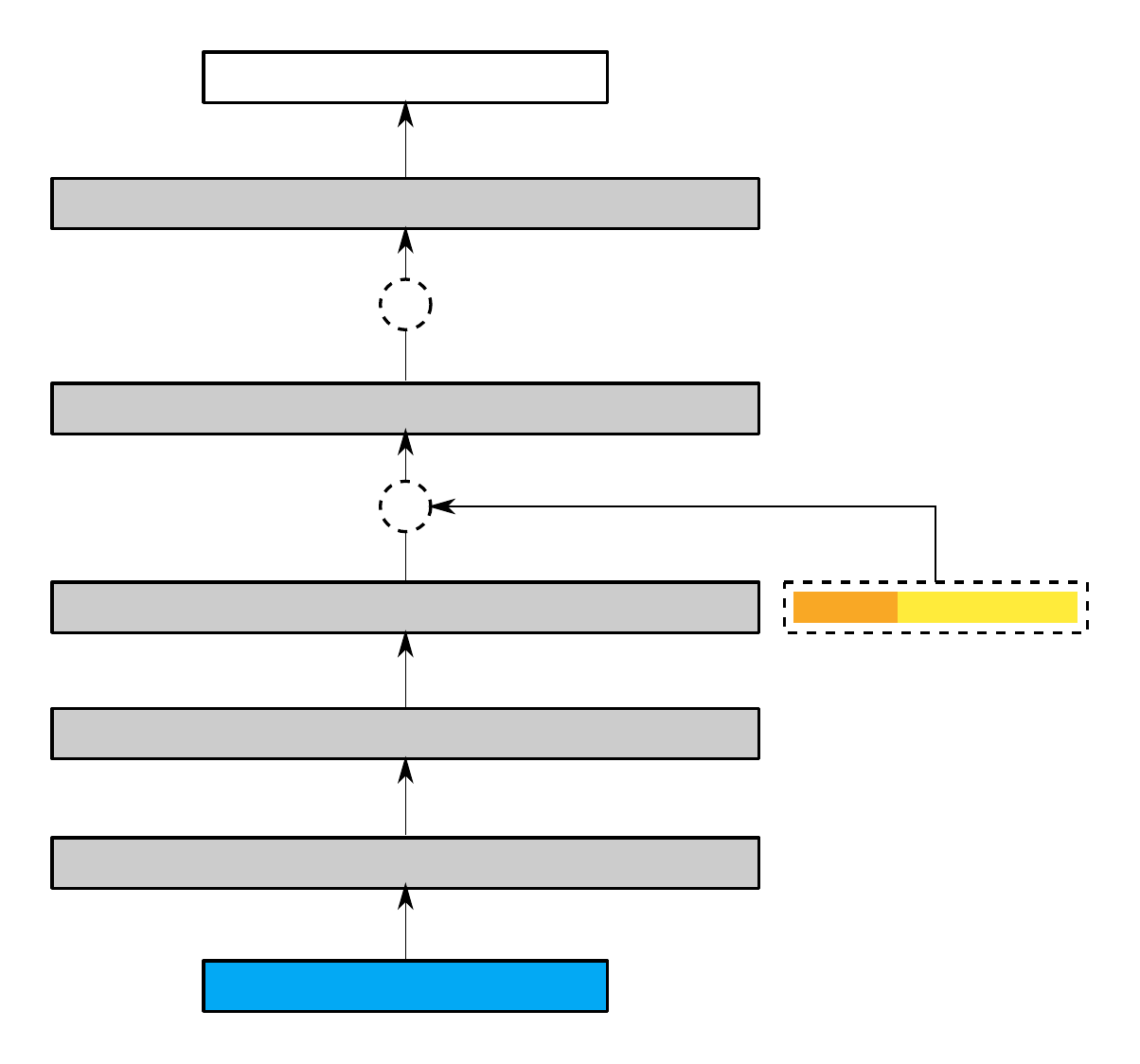} & 
  \includegraphics[width=0.45\columnwidth]{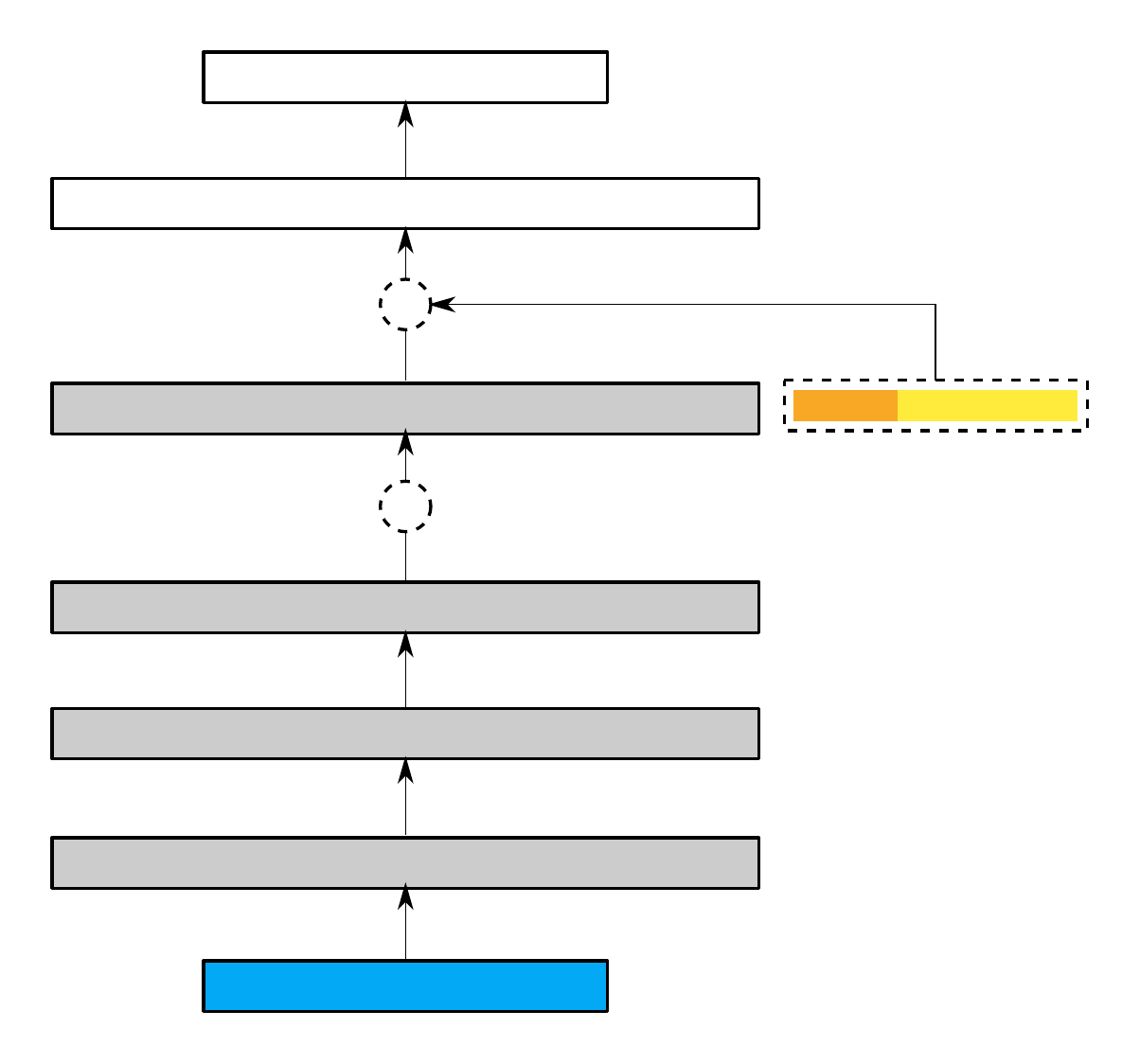} \\
  (a) Nonlinear case & (b) Linear case
  \end{tabular}
  \vspace{-5mm}
\caption{Different injection points of proposed factorized speaker transformation. It may be applied to intermediate hidden layers with non-linear activation functions or used at a specific layer where all remaining operations are linear. Relationships between speaker transforms and acoustic features are non-linear for the former case but linear for the latter case.}
\label{fig:activation}
\vspace{-3mm}
\end{figure}

The second strategy is to combine the proposed method with other type of matrix decomposition. For example, in the work of Xue et al. \cite{xue2013restructuring}, a weight matrix is decomposed into three linearly connected matrices using singular value decomposition (SVD). Therefore, instead of multiplying a scaling matrix to a weight matrix, we may first decompose the weight matrix into the three linearly connected matrices and use the proposed scaling matrix to approximate one of the decomposed matrices further as follows: 
\begin{align}
\label{eq:svdlayer}
    \mathbf{\overline{h}}_{l} &= f(\mathbf{W}^{(k)}_l\mathbf{h}_{l-1} + \mathbf{c}_l + \mathbf{b}^{(k)}_{l} +  \mathbf{h}_{l-1})\\
    \mathbf{W}^{(k)}_l &= \mathbf{U}_{l}\mathbf{A}^{(k)}_l\mathbf{V}_{l}\\
    \mathbf{A}^{(k)}_{l} &= {\rm diag} (\mathbf{W}^{A}_l\mathbf{s}^{A,(k)}) \\
    \mathbf{b}^{(k)}_{l} &= \mathbf{W}^{b}_l\mathbf{s}^{b,(k)}
\end{align}
where $\mathbf{U}_l \in \mathbb{R}^{m \times n}$, $\mathbf{V}_l \in \mathbb{R}^{n \times m}$ and $\mathbf{A}^{(k)}_l \in \mathbb{R}^{n \times n}$ with $n \ll m$\footnote{It is also possible to theoretically include SVD bottleneck speaker adaptation with low-rank approximation \cite{xue2014singular}. To do this, a constrain $\mathbf{W}_l \approx \mathbf{U}_{l}\mathbf{V}_{l}$ needs to be added.}. Note that residual connections are also added here. When we use this model for time-series speech data, the input varies at each time and the residual part becomes a time-variant bias term as $\mathbf{\overline{h}}_{l,t} = f(\mathbf{W}^{(k)}_l\mathbf{h}_{l-1,t} + \mathbf{c}_l + \mathbf{b}^{(k)}_{l} +  \mathbf{h}_{l-1,t})$ where $\mathbf{h}_{l,t}$ is output of the $l$-th hidden unit at time $t$. The bottleneck method can be summarized as Figure \ref{fig:strategies}-b. 



\begin{table*}[t]
    \caption{Divisions of English and Japanese speech corpora used in our experiments.}
    \centering
    \scalebox{0.8}{
    \begin{tabular}{|l|r|r|r|r|r|r|r|r|r|}
        \hline
             \multirow{2}{*}{Set} & \multicolumn{2}{c|}{Train (Speech \& Text)} & \multicolumn{2}{c|}{Valid (Speech \& Text)} & \multicolumn{2}{c|}{Test (Text)} & \multicolumn{3}{c|}{Speakers}\\ \cline{2-10}
           & Each speaker & Total & Each speaker & Total & Each speaker & Total &Male & Female & Total  \\ \hline \hline
         en.base & $\sim$370 & 26785 & 5 & 360 & - & - & 31 & 41 & 72 \\ \hline
         en.target.10 & 10 & 80 & \multirow{4}{*}{5} & \multirow{4}{*}{40} & \multirow{4}{*}{15} & \multirow{4}{*}{120} &  \multirow{4}{*}{4} & \multirow{4}{*}{4} & \multirow{4}{*}{8}\\
         en.target.40 & 40 & 500 & & & & & & & \\
         en.target.160 & 160 & 1280 & & & & & & & \\ 
         en.target.320 & 320 & 2560 & & & & & & &\\ \hline \hline
         jp.base & $\sim$148 & 34713 & 3 & 705 & - & - & 51 &	184 & 235 \\ \hline
         jp.target.10 & 10 & 200 & \multirow{3}{*}{3} & \multirow{3}{*}{60} & \multirow{3}{*}{10} & \multirow{3}{*}{200} & \multirow{3}{*}{10} & \multirow{3}{*}{10} & \multirow{3}{*}{20}\\
         jp.target.50 & 50 & 1000 & & & & & & & \\
         jp.target.100 & 100 & 2000 & & & & & & & \\ \hline

    \end{tabular}}
    \label{tab:dataset}
  \vspace{-5mm}
\end{table*}

\begin{table}[tb]
    \caption{Different strategies evaluated in this paper. The parameter's size was purposely chosen so that all models used the same number of parameters. }

    \centering
    \scalebox{0.8}{
    \begin{tabular}{llrrr}
        \hline
             & & \multicolumn{3}{c}{Size} \\ \cline{3-5}
         Notation  & Strategy & Scaling & Bias & Bottleneck  \\ \hline
         bias     & bias code      & -  & 64 & -\\
         scale    & scaling code   & 64 & -  & - \\
         affine   & bias + scaling & 32 & 32 & - \\
         level    & multilevel     & 32 & 32 & - \\
         bottle   & bottleneck & 64 & 32 & 512  \\ \hline
    \end{tabular}}
    \label{tab:models}
  \vspace{-3mm}
\end{table}
We also investigate to which layers we should inject the proposed transformation and what kinds of activation functions should be used after the speaker transformation. More specifically, we investigate whether the proposed transformation should be used at intermediate hidden layers with non-linear activation functions as shown in Figure \ref{fig:activation}-a or at a specific layer where all remaining operations are linear as shown in Figure \ref{fig:activation}-b. By analyzing this, we can understand whether the relationship between the proposed speaker transformation functions and generated acoustic features should be represented in a non-linear way like the former case, or in a linear one like the latter case.\footnote{For the combination of the linear case with the strategy in Figure \ref{fig:strategies}-a, which has operations at two different layers, we first used speaker transformation based on the bias code at a hidden layer with the non-linear activation functions and further used speaker transformation based on the scaling code at the next linear layer. This is technically a mix of linear and non-linear speaker transformations, but we included this in "the linear setup" in our experiments.}




\section{Experiments}
\label{sec:exp}

\subsection{Experimental condition}
\label{subsec:condition}

We use two speech corpora to evaluate our proposal: an English corpus containing 80 speakers, which is a subset of the VCTK \cite{veaux2013voice,veaux2017superseded}, and an in-house Japanese speech corpus with over 250 speakers. The English corpus was used to objectively evaluate various aspects of our proposal while the Japanese corpus is used to reproduce the results and evaluate subjectively with native Japanese listeners.
We split each corpora into the base and target sets as shown in Table \ref{tab:dataset} and conducted two tasks (multi-speaker and adaptation) as follows. In the multi-speaker task, we used en.base and one of en.target.\{10, 40, 160, or 320\} for training a multi-speaker neural network common to all speakers per strategy. In the adaptation task, we used en.base for training a multi-speaker neural network per strategy and adapted it to each target speaker included in en.target.*. In both the tasks, the evaluation was performed using target speakers included in en.target.*. This increased the number of models needed to be constructed but reduced the mismatch between the multi-speaker and adaptation tasks so we could directly compare them. 

\begin{figure}[tb]
  \centering
  \includegraphics[width=0.85\columnwidth]{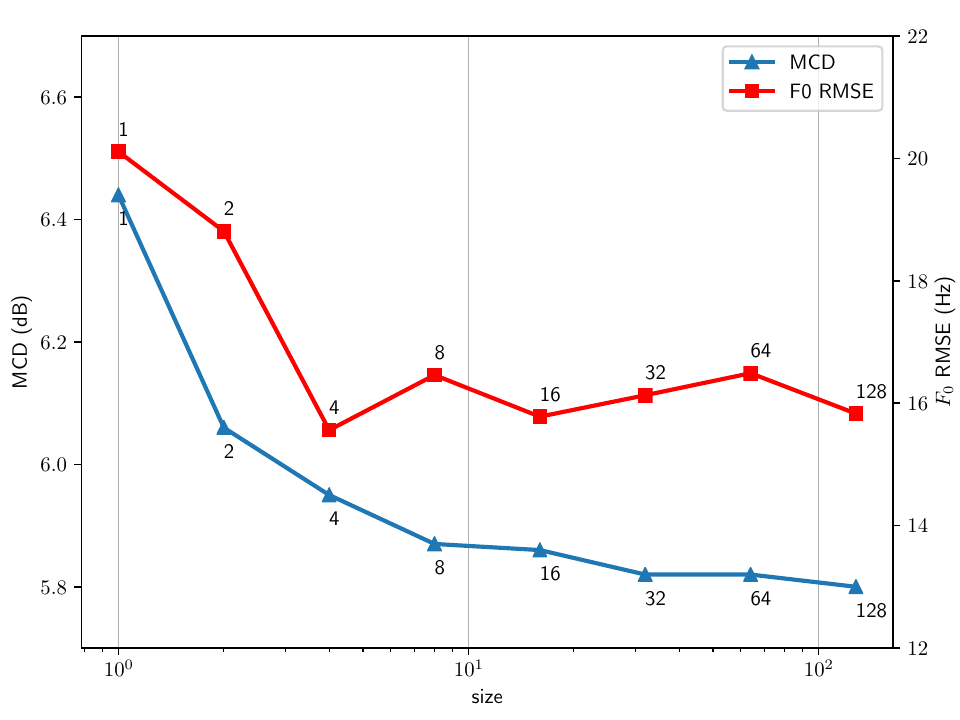}
  \vspace{-5mm}
\caption{Objective evaluation of changing size of scaling code in nonlinear setup.}
\label{fig:scale_size}
\vspace{-6mm}
\end{figure}

For the DNN-based acoustic model, we used a conventional multi-task learning neural network similar to our previous works \cite{luong2017adapting,luong2018unsupervised}. The neural network maps linguistic features (depending on languages) to several acoustic features including 60-dimensional mel-cepstral coefficients, 25-dimensional band-limited aperiodicities, interpolated logarithm fundamental frequencies, and their dynamic counterpart. A voiced/unvoiced binary flag is also included. 
The neural network model has five feedforward layers each with 1024 neurons, followed by a linear layer to map to the desired dimensional output. All layers have the sigmoid activation function unless stated otherwise. We experimented with five strategies utilizing either scaling code, bias code, or both as shown in Table \ref{tab:models}. Further, to investigate the impacts of different waveform generation methods, we used both a speaker-independent Wavenet vocoder \cite{van2016wavenet,hayashi2017investigation} and the WORLD vocoder \cite{morise2016world} for speech waveform generation . However, our Wavenet model is still under development and we experienced the collapse of generated speech problems, which is described in \cite{wu2018collapsed}. 

\vspace{-2mm}
\subsection{Objective evaluation}
\label{subsec:obj}

\begin{figure}[tb]
  \centering
  \includegraphics[width=1.0\columnwidth]{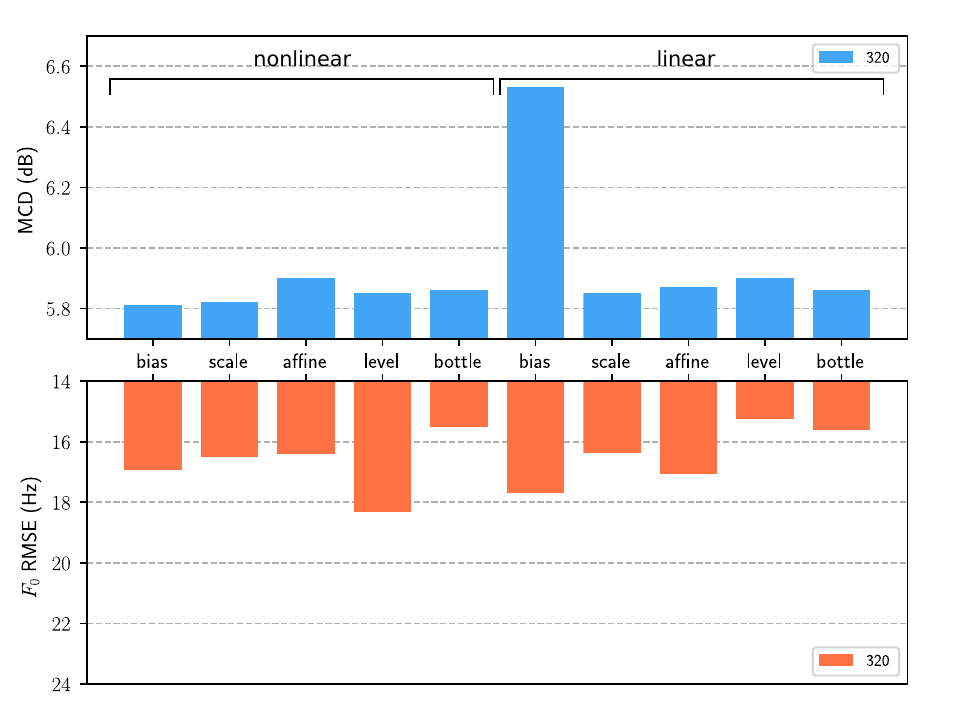}
  \vspace{-8mm}
   \caption{Objective evaluations results of different strategies in the multi-speaker task using the English corpus.}
\label{fig:obj_multispeaker}
\vspace{-5mm}
\end{figure}

We first evaluated the scaling code by itself in a nonlinear setup since, at the time of writing, using scaling code for multi-speaker speech synthesis has not been investigated. We changed the size of scaling codes from 1 to 128 to see how they impact the objective performance of the multi-speaker task in a similar way to experiments that we did on bias codes previously \cite{luong2017adapting}. The multi-speaker models were trained using en.base and en.target.320 together. 
The objective evaluation results, including mel-cepstral distortion (MCD) in dB and $F_0$ root mean square error ($F_0$ RMSE) in Hz, are illustrated in Figure \ref{fig:scale_size}.
We can see that both the distortions decrease when we increase the size of the scaling code. 

Next we evaluated multiple strategies described in Table \ref{tab:models} for the multi-speaker task in either nonlinear or linear setups. Again the multi-speaker models were trained using the en.base and en.target.320 data together. Figure \ref{fig:obj_multispeaker} shows objective evaluation results of the strategies. If we look at the non-linear setups, we see that there are no obvious differences between these strategies. However, at least we can determine that the proposed scaling code can be used by itself without decreasing the performance. If we look at the linear setups, we can clearly see that the using the bias code by itself is a poor strategy for multi-speaker modeling. It resulted in much worse MCD even though its $F_0$ RMSE is comparable to other systems. In \cite{wang2018investigating}, Wang found out that the model structures required for mel-cepstrum and fundamental frequency are different. Our results also support this finding.

\begin{figure}[tb]
  \centering
  \includegraphics[width=1.0\columnwidth]{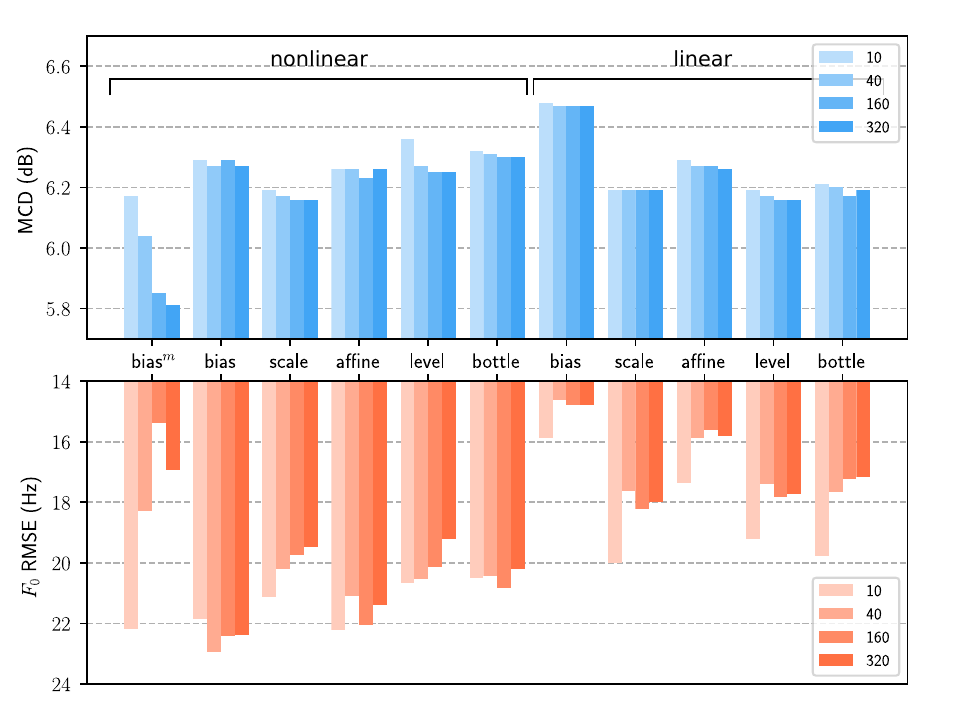}
  \vspace{-7mm}
\caption{Objective evaluation results of different strategies in adaptation task using English corpus. Here bias$^{m}$ shows  reference results in the multi-speaker task using the bias code in the nonlinear setup. All other results are for adaptation of unseen speakers using data included in en.target.*.}
\label{fig:obj_adaptation}
\vspace{-5mm}
\end{figure}

Figure \ref{fig:obj_adaptation} shows objective evaluation results of the strategies in the adaptation task using different amounts of data. The first block indicated bias$^{m}$ corresponds to reference results in the multi-speaker task (i.e.,\ systems where multi-speaker neural networks were trained using en.base and one of en.target.\{10, 40, 160, or 320\} and synthetic speech was generated using text of the test set of target speakers ) using the bias code in the nonlinear setup. All other results are adaptation results for the unseen speaker task. The amounts of adaptation data vary from 10 to 320.

From this figure, we see that adaptation to the unseen speakers is more difficult than multi-speaker modeling. Moreover, while the results of multi-speaker modeling are improved significantly when we increase the amount of data, the adaptation results for the unseen speakers show marginal improvements when more data is available. 
This suggests that the proposed adaptation transformation needs to be generalized better. 
Another important pattern that we can see from the figure is that in terms of $F_0$ RMSE, all strategies in the linear setup outperform their nonlinear counterparts. 

\vspace{-2mm}
\subsection{Subjective evaluations}
\label{subsec:subjective}

\begin{figure}[tb]
  \centering
  \includegraphics[width=1.0\columnwidth]{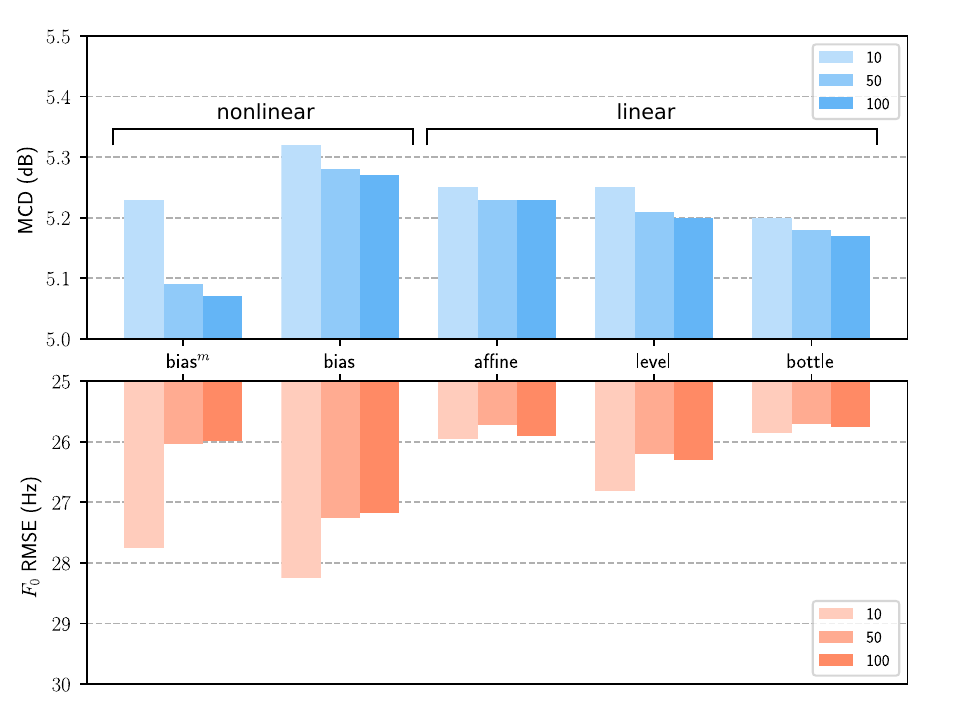}
  \vspace{-7mm}
\caption{Objective evaluation results of selected strategies in adaptation task using Japanese corpus. Like the English test,  bias$^{m}$ shows reference results in the multi-speaker task using the bias code in the nonlinear setup. All other results are adaptation results.}
\label{fig:obj_kanto}
\vspace{-5mm}
\end{figure}

Next we reproduced several selected strategies using the Japanese dataset. We doubled the size of speaker codes shown in Table \ref{tab:models} and chose strategies that showed reasonable improvements in the objective evaluation using the English dataset. The objective evaluation results using the Japanese corpus are shown in Figure \ref{fig:obj_kanto}, from which we can see the same trend as the result using the English one\footnote{Speech samples using the English corpus can be found at \url{http://www.hieuthi.com/papers/slt2018}}. 

We used the Japanese systems and conducted a subjective listening test to see how participants perceived these differences. The listening test contained two sets of questions. In the first part, participants were asked to judge the naturalness of the presented speech sample using a five-point scale ranged from 1 (very unnatural) to 5 (very natural). In the second part, participants were asked to compare a speech sample of a system with recorded speech of the same speaker and judge if they are the same speaker or not using a four-point scale ranged from 1 (different, sure) to 4 (same, sure). This evaluation methodology is similar to our previous study \cite{luong2018unsupervised}. In addition to synthetic speech generated from the proposed speech synthesis systems using the above selected strategies, we also evaluated recorded speech, WOLRD vocoded speech, and Wavenet vocoded speech for comparison. A large-scale listening test was done with 289 subjects. The statistical analysis was conducted using pairwise t-tests with a 95\% confidence margin and Holm-Bonferroni compensation for multiple comparisons. 

\begin{figure}[tb]
  \centering
  \includegraphics[width=1.0\columnwidth]{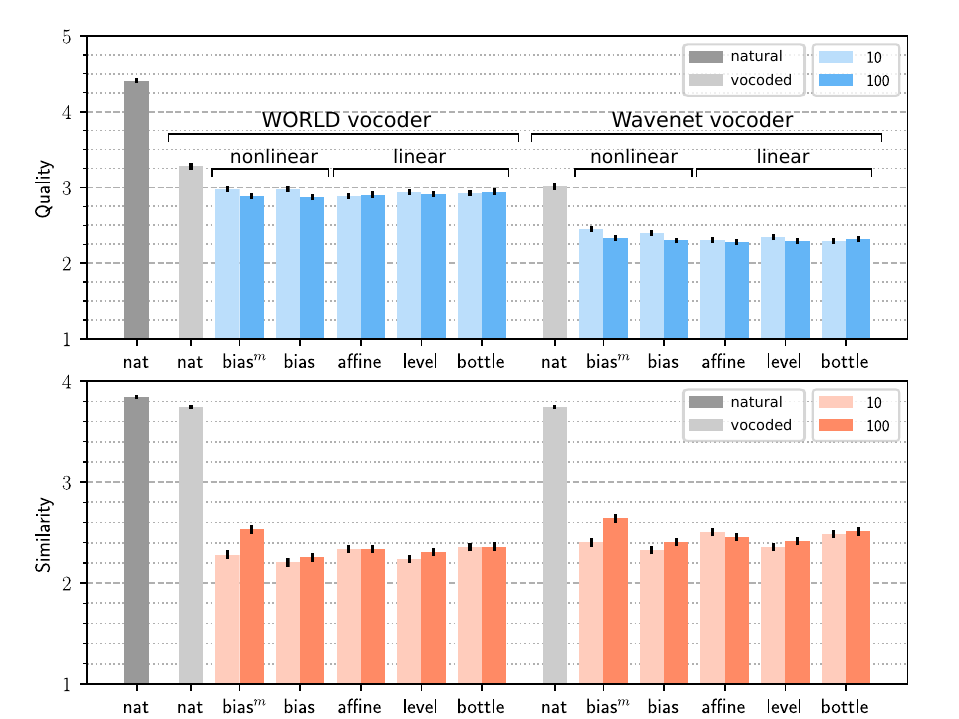}
  \vspace{-7mm}
\caption{Subjective evaluation results of selected strategies in adaptation task using Japanese corpus. 
Top figure shows mean opinion scores on naturalness. Bottom figure shows speaker similarity scores. Recorded speech and vocoded speech using correct acoustic features were also evaluated at the same time.}
\label{fig:sbj_kanto}
\vspace{-3mm}
\end{figure}

Subjective evaluation results are presented in Figure \ref{fig:sbj_kanto}. In the quality test, we can first see that participants judged all systems using our speaker-independent Wavenet vocoder samples to be worse than counterparts using the WORLD vocoder. This is inconsistent with other publication results and indicates that our Wavenet is not properly trained. For the future works, we could further fine-tune a part of the speaker-independent Wavenet model to stabilize the neural-net vocoder \cite{ustc018wavenet-adaptation,sisman2018voice}.  However, unlike the quality test, the subjects judged synthetic speech using the Wavenet vocoder to be closer to the target speakers in the speaker similarity test although there are still large gaps between vocoded speech and synthetic speech.

We can also see that a reference multi-speaker system marked as bias$^{m}$ using 100 utterances has the highest similarity score among the other systems, and this is consistent with the objective evaluation results. Regarding the adaptation to the unseen speakers, we could see that the proposed method using both the scaling and bias codes and its bottleneck variant (in the linear setting) have better results than the adaptation method using the bias code in the nonlinear setting (which is our previous work) for both WORLD and Wavenet vocoders. This would be because of improved F0 adaptation, as we can see objectively in Figure \ref{fig:obj_kanto}. Regarding the quantity of the adaptation data, more data seems to slightly improve speaker similarity of synthetic speech in general but does not improve the perception of quality. In some cases, it makes the quality of synthetic speech slightly worse.  





\section{Conclusions}
\label{sec:conclusions}
\vspace{-2mm}


In this paper, we have explained several major existing adaptation frameworks for DNN speech synthesis and showed one generalized speaker-adaptive transformation. Further, we have factorized the proposed transformation on the basic of scaling and bias codes and investigated its variants such as bottleneck. 

From objective and subjective experiments, we showed that the proposed method, specifically the ones using both the scaling and bias codes in the linear setting, can reduce acoustic errors and improve subjective speaker similarity in the adaptation of unseen speakers . 
Moreover, our results clearly indicate that there are still large gaps between vocoded speech and synthetic speech in terms of speaker similarity and this clearly indicates that there is room for improving multi-speaker modeling and speaker adaptation. 

Our future work includes comparing our method with other adaptation methods such as LHUC and SVD bottleneck speaker adaptation with low-rank approximation. Another interesting experiment we would like to see is the use of i-vector or d-vector \cite{doddipatla2017speaker} as a scaling code.

\bibliographystyle{IEEEbib}
\bibliography{main}

\end{document}